\documentclass[12pt]{ronbun}

\usepackage{amsmath,amssymb}
\usepackage{graphicx,color}
\usepackage{cite}
\usepackage{bm}
\usepackage{dcolumn}

\newcommand{\nn}{\nonumber}
\newcommand{\e}{{\rm e}}

\renewcommand{\th}{\theta}

\numberwithin{equation}{section}

\begin{document}

\begin{flushright}
\parbox{4.2cm}
{KEK-TH-1009 \hfill \\
{\tt hep-th/0503057}
 }
\end{flushright}

\vspace*{2cm}

\begin{center}
 \Large\bf 
Bubbling 1/2 BPS Geometries and Penrose Limits  
\end{center}
\vspace*{1.2cm}
\begin{center}
{\large Yastoshi Takayama$^{\ast\dagger a}$ 
and Kentaroh Yoshida$^{\ast b}$}
\end{center}
\vspace*{0.5cm}
\begin{center}
$^{\ast}$\emph{Theory Division,   	
Institute of Particle and Nuclear Studies, \\
High Energy Accelerator Research 
Organization (KEK),\\
Tsukuba, Ibaraki 305-0801, Japan.} \\
$^{\dagger}$\emph{
Department of Particle and Nuclear Physics, \\
The Graduate University for Advanced Studies, \\
Tsukuba, Ibaraki 305-0801, Japan.} 

\vspace*{0.3cm}
$^a${\tt takaya@post.kek.jp} \quad 
$^b${\tt kyoshida@post.kek.jp}
\end{center}

\vspace*{1.5cm}

\centerline{\bf Abstract}

\vspace*{0.5cm}
We discuss how to take a Penrose limit in bubbling 1/2 BPS geometries at the stage of a single function $z(x_1,x_2,y)$\,. 
By starting from the $z$ of the AdS$_5\times$S$^5$ we can directly derive that of the pp-wave via the Penrose limit.
In course of the calculation the function $z$ for the pp-wave with $1/R^2$-corrections is obtained. 
We see that it surely reproduces the pp-wave with $1/R^2$ terms. 
We also investigate the pp-wave with higher order $1/R^2$-corrections.
In addition the Penrose limit in the configuration of the concentric rings is considered. 

\vspace*{0.5cm}

\vfill \noindent {\bf Keywords:}~~{\footnotesize Bubbling, pp-wave,
Penrose limit} 

\thispagestyle{empty}
\setcounter{page}{0}

\newpage

\section{Introduction}
Recently we have a renewed interest in the AdS/CFT duality \cite{M} in
the 1/2 BPS sector. Kaluza-Klein gravitons, giant gravitons \cite{giant}
and dual giant gravitons \cite{dual}, who correspond to 1/2 BPS states
in the Super Yang-Mills (SYM) side, are well described by using the free fermion
description \cite{Jevicki}. An application of the free fermion
description to study the AdS/CFT duality is recently discussed by
Berenstein \cite{Berenstein}. (The free fermion description of giant
gravitons in this direction is further studied in \cite{Silva}.) The
supergravity description of the phase space of the free fermion is
clarified by Lin-Lunin-Maldacena (LLM) \cite{LLM}. The 1/2 BPS solutions
of type IIB supergravity preserving an isometry $R\times SO(4)\times
SO(4)$ are characterized by a single function satisfying a differential
equation. The function can be obtained by giving a boundary condition in
a two-dimensional subspace in ten-dimensional spacetime.  The phase
space of the free fermion may be identified with the two-dimensional
plane in which droplets are drawn for each of boundary conditions
\cite{LLM}.  The Pauli exclusion principle in the free fermion is
intimately related to the causality in the supergravity solutions
\cite{Pauli}. The single function parameterizing the 1/2 BPS geometries
is also related to the Wigner phase-space distribution \cite{Mandal}.
In addition, topological transitions in bubbling 1/2 BPS geometries are
also discussed in \cite{Horava}.

The description of bubbling 1/2 BPS geometries is extended in various
directions. The generalizations to other dimensions or backgrounds are
considered in \cite{other-dim}. Some semiclassical strings \cite{GKP2}
on the geometries for the configuration of concentric rings \cite{LLM}
are studied in \cite{FJ,EM}. The tiny giant graviton matrix approach is
considered in \cite{tiny}. The extension to the finite temperature case
is discussed in \cite{finite}. 

In this paper we discuss how to take a Penrose limit \cite{P} in bubbling 1/2 BPS geometries at the stage of function $z(x_1,x_2,y)$\,.
As discussed in \cite {LLM}, the Penrose limit is interpreted as the magnification of a part of the droplet. 
This interpretation is quite natural since the Penrose limit implies the magnification around a certain null geodesics.
Following LLM's observation, we directly show that the function $z$ for the  AdS$_5\times$S$^5$ is reduced to the one for the pp-wave \cite{BFHP1} via the Penrose limit.
In course of the calculation we obtain the $z$ for the pp-wave with  $1/R^2$-corrections as a byproduct.
This $z$ surely gives the pp-wave metric with $1/R^2$ terms discussed in \cite{Schwarz}.
It should be noted that this $z$ has the same boundary as the pp-wave without $1/R^2$ corrections at the $1/R^2$ order level.
This result implies a subtlety to take account of $1/R^2$-corrections at the level of a single function $z$\,, and so it seems difficult to obtain the function $z$ with $1/R^2$-corrections by directly carrying out the integral for $z$ under a boundary condition. 
We also investigate higher order $1/R^2$-corrections.
It is found that the higher order corrections do {\it not} modify the half-filling configuration and such a subtlety is not improved.
Moreover we consider the Penrose limit in the geometries for the configuration of the concentric rings.

This paper is organized as follows:
In section 2 we briefly introduce 1/2 BPS geometries obtained by LLM.
In section 3 we discuss how to take a Penrose limit in bubbling 1/2 BPS geometries and obtain the single function $z$ with $1/R^2$-corrections.
In section 4 the higher order $1/R^2$-corrections are investigated.
In section 5 the result in the section 3 is applied to the concentric ring case.
Section 6 is devoted to a conclusion and discussions.

\section{Setup}

\begin{figure}
 \begin{center}
  \includegraphics[scale=.8]{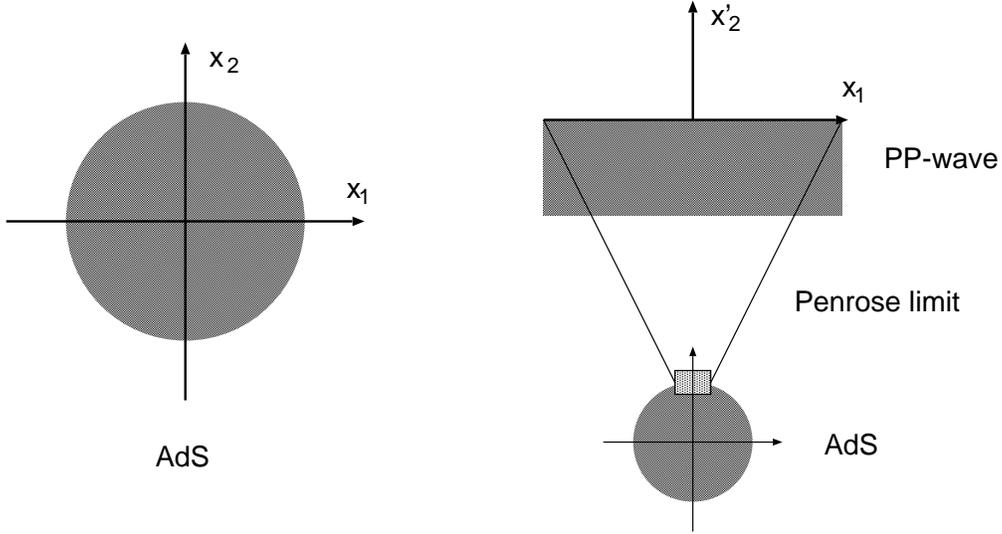} 
 \end{center}
\caption{AdS, pp-wave and Penrose limit.}
\label{pp-wave:fig} 
\end{figure}

All 1/2 BPS geometries of type IIB supergravity preserving the
isometry $R\times SO(4)\times SO(4)$ are obtained by Lin-Lunin-Maldacena
\cite{LLM}. The 1/2 BPS geometries are given by
\begin{eqnarray}
&& ds^2 = -h^{-2}\left(dt + \sum_{i=1}^2V_{i}dx^i\right)^2 
+ h^2 \left(dy^2 + \sum_{i=1}^2dx^idx^i\right) + y\e^{G}d\Omega^2_3 + y\e^{-G}d\widetilde{\Omega}^2_3\,, \label{metric} \qquad \\
&& \qquad 
 h^{-2} = 2y\cosh G\,, \quad z\equiv \frac{1}{2}\tanh G\,, \quad 
\e^{G} = \sqrt{\frac{1+\tilde{z}}{-\tilde{z}}}\,, 
\quad
\tilde{z} \equiv z  - \frac{1}{2}\,,  
\nn \\ 
&& \qquad y\partial_yV_i = \epsilon_{ij}\partial_j z\,, \quad 
y(\partial_iV_j-\partial_jV_i) = \epsilon_{ij}\partial_yz\,, \label{z-v} \\ 
&& \qquad 
B_t = - \frac{1}{4}y^2\e^{2G}\,, \quad \widetilde{B}_t 
= - \frac{1}{4}y^2\e^{-2G}\,,\nn \\ 
&& \qquad F = dB_t\wedge (dt+V)+B_tdV + d\widehat{B}\,, \quad 
 \widetilde{F} = d\widetilde{B}_t\wedge(dt+V) + \widetilde{B}_t dV 
+d\widehat{\widetilde{B}}\,, \nn \\
&& \qquad  
d\widehat{B} = -\frac{1}{4}y^3\ast_3d 
\left(\frac{z+\frac{1}{2}}{y^2}\right)\,, 
\quad d\widehat{\widetilde{B}} = - \frac{1}{4}y^3\ast_3 d\left(
\frac{z-\frac{1}{2}}{y^2}\right)\,, \nn
\end{eqnarray}
where a single function $z(x^1,x^2,y)$ satisfies the following
differential equation: 
\begin{eqnarray}
\partial_i\partial_i z + y\partial_y\left(\frac{\partial_y z}{y}\right) = 0\,. 
\label{diff-eq} 
\end{eqnarray} 
Remarkably, the single function $z$ determines the solution of type IIB
supergravity preserving an isometry $R\times SO(4)\times SO(4)$\,. If
one would impose an appropriate boundary condition, then one can solve
the differential equation and obtain the solution $z$\,. That is, when
we give a boundary condition the solution of the supergravity is
determined. The possible boundary conditions are severely restricted by
requiring the smoothness of the solution.
This requiring allows the function $z$ to take two values $z= \pm 1/2$ at $y=0$.
When we assign white
and black to $z=1/2$ and $z=-1/2$\,, respectively, the droplet
configurations can be drawn in the $(x_1,x_2)$-plane. This plane is
identified with the phase space of the free fermion discussed by
Berenstein \cite{Berenstein}. In particular, the configuration of a
single black disk corresponds to the AdS$_5\times$S$^5$ case and the
configuration that lower half-plane is filled describes the pp-wave
background. Then the Penrose limit is interpreted as the magnification
of a part of the geometry for the AdS. These are depicted in
Fig.\,\ref{pp-wave:fig}.

\section{Penrose Limit of $z$ for AdS$_5\times$S$^5$}
\label{sec3}

To begin with, we shall consider the Penrose limit of the
function $z$ for the AdS$_5\times$S$^5$ case:  
\begin{eqnarray}
\tilde{z}(r,y;r_0) 
= \frac{r^2-r^2_0+y^2}{2\sqrt{(r^2+r^2_0+y^2)^2-4r^2r^2_0}} -\frac{1}{2}\,.  
\end{eqnarray}  
The coordinate system of the LLM background for this $z$ is different
from the standard global coordinates of the AdS$_5\times$S$^5$ background. 
In order to obtain the standard expression of the AdS$_5\times$S$^5$ 
we need to perform the change of coordinates as follows: 
\begin{eqnarray}
y = r_0\sinh\rho\,\sin\theta\,, \quad 
r = r_0\cosh \rho\,\cos\theta\,, \quad \widetilde{\phi} = \phi -t\,. 
\end{eqnarray}
Here the radius $r_0$ is identified with the AdS radius $R$ via $r_0 =
R^2$\,. 

Let us recall how to take the Penrose limit in the metric. 
The AdS$_5\times$S$^5$ metric with global coordinates is given by 
\begin{eqnarray}
ds^2 = R^2 \left[-\cosh^2\rho \, dt^2 + d\rho^2 + \sinh^2\, \rho d\Omega_3^2 
+ \cos^2\th\, d\widetilde{\phi}^2 + d\th^2 +\sin^2\th\, d\widetilde{\Omega}_3^2\right]\,.
\end{eqnarray}
Here we introduce the following parameterization utilized by Callan et.al
\cite{Schwarz}: 
\begin{eqnarray}
\cosh\rho = \frac{1 + r_1^2/4}{1 - r_1^2/4}\,, 
\qquad \cos\th = \frac{1-r^2_2/4}{1+r^2_2/4}\,. 
\end{eqnarray} 
When we will take a limit $R\rightarrow\infty$ by rescaling $r_1$ and
$r_2$ as $r_1 \rightarrow r_1/R$ and $r_2 \rightarrow r_2/R$,
respectively, the resulting function $z$\,, up to and including $1/R^2$\,, is  
\begin{eqnarray}
z = \frac{r_1^2 -r_2^2}{2(r_1^2+r_2^2)} + \frac{r_1^2r_2^2}{2(r_1^2+r_2^2)}
\frac{1}{R^2} + \mathcal{O}(1/R^4)\,. 
\label{z-ex} 
\end{eqnarray}
In order to obtain the function $z$ in terms of $x_1,x_2,y$\,, 
we need to perform a coordinate transformation from $(r_1,r_2)$ to
$(r,y)$\,. In course of the calculation $1/R^2$ corrections appear and so we have to
be careful to do it. 

We firstly expand $y$ and $r$ with respect to $r_1$ and $r_2$ (after the
rescaling) as   
\begin{eqnarray}
\label{eq1}
&& y = r_1 r_2 + \frac{1}{4R^2}r_1r_2(r_1^2 - r_2^2) + \mathcal{O}(1/R^4)
\,, \\ 
&& r = r_0 + \frac{1}{2}(r_1^2 - r^2_2) + \frac{r_0}{8R^2}(r_1^2-r_2^2)^2 
+ \mathcal{O}(1/R^4)\,.  \label{eq2}
\end{eqnarray}
From (\ref{eq1}) and (\ref{eq2}) we obtain
\begin{eqnarray}
\frac{1}{2R^2}(r_1^2 - r_2^2) = -1\pm \sqrt{\frac{2r}{r_0}-1}\,, \qquad 
r_1r_2 = \frac{2y}{1\pm \sqrt{\frac{2r}{r_0}-1}}\,. \label{rs}
\end{eqnarray} 
We have two choices to take a pair of $r_1$ and $r_2$ due to the sign
$\pm$ that appears when we solved the quadratic algebraic equation
(\ref{eq2})\,.  But we should choose the ``$+$'' sign that leads to
$y=r_1r_2$ at the zero-th order of $1/R^2$\,. The relation $y=r_1r_2$ is
utilized in \cite{LLM}. By using (\ref{rs}) we can rewrite the first
term in (\ref{z-ex}) as 
\begin{eqnarray}
\frac{r_1^2 - r^2_2}{2(r_1^2+r_2^2)} = \frac{r-r_0}{
2\sqrt{\left(r-r_0 \right)^2 + y^2}}\,. 
\end{eqnarray}
Now let us introduce a new variable $x_2' = x_2 - r_0$\,. This shift of
$x_2$ corresponds to that of the origin in the 2-plane. That is, the
origin in the system of coordinates 
for the AdS
is shifted to the north-pole of the disk, and the resulting origin is
nothing but the origin in 2-plane for the pp-wave (see
Fig.\,\ref{pp-wave:fig}).  Then it is possible to expand $r^2 = x_1^2 +
x_2^2$ as
\begin{eqnarray}
r = \sqrt{x_1^2 + (x_2'+r_0)^2} = (x_2' + r_0) 
+ \frac{x_1^2}{2(x_2'+r_0)} + \cdots\,. 
\label{expan}
\end{eqnarray}
By using (\ref{expan}), we can express the first term 
in (\ref{z-ex}) in terms of $(x_1,x'_2)$: 
\begin{eqnarray}
\frac{r_1^2-r_2^2}{2(r_1^2+r_2^2)} = \frac{x_2'}{2\sqrt{x_2'{}^2+y^2}} + 
\frac{x_1^2\,y^2}{4 R^2 (x_2'{}^2+y^2)^{3/2}} +\mathcal{O}(1/R^4)\,.  
\end{eqnarray}
Thus the first term in (\ref{z-ex}) has been decomposed into 
the leading term and the sub-leading term. As a matter of course, the
leading term agrees with the result of LLM \cite{LLM}. 
 
In order to finish evaluating the $1/R^2$-corrections it is necessary to
investigate the second term in (\ref{z-ex})\,. It is easy to show that
\begin{eqnarray}
\frac{r_1^2r_2^2}{2R^2(r_1^2 + r_2^2)} = \frac{y^2}{4R^2\sqrt{x_2'{}^2+y^2}}
+ \mathcal{O}(1/R^4)\,, 
\end{eqnarray} 
and so the resulting function $z$ with $1/R^2$ contributions is given by
\begin{eqnarray}
&& z(x_1,x_2',y) 
= \frac{r_1^2 -r_2^2}{2(r_1^2+r_2^2)} + \frac{r_1^2r_2^2}{2(r_1^2+r_2^2)}
\frac{1}{R^2} + \mathcal{O}(1/R^4) \nn \\ 
&& \qquad \qquad~~ = \frac{x_2'}{2\sqrt{x_2'{}^2 +y^2}} 
+ \frac{1}{4R^2}\left[
 \frac{x_1^2\,y^2}{(x_2'{}^2+y^2)^{3/2}} + \frac{y^2}{\sqrt{x_2'{}^2+y^2}}
\right] + \mathcal{O}(1/R^4)\,. \label{z}
\end{eqnarray} 
It is an easy task to show directly that the function $z$ given in
(\ref{z}) satisfies the differential equation (\ref{diff-eq})\,. Hence
the pp-wave with $1/R^2$ terms may be contained in the context of
bubbling 1/2 BPS geometries. 

We shall next consider the Penrose limit of the $V_r$ and $V_{\phi}$ 
for the AdS: 
\begin{eqnarray}
V_r =0\,, \qquad V_{\phi} = -\frac{1}{2}\left(
\frac{r^2 + r_0^2 + y^2}{\sqrt{(r^2+r^2_0+y^2)^2-4r^2r^2_0}}-1\right)\,. 
\end{eqnarray}
For the pp-wave case the Cartesian coordinates $(x_1,x_2)$ are more
suitable than the polar coordinates $(r,\phi)$ since the droplet
configuration is the lower-half plane rather than a disk.
Through the coordinate transformations we can find the following $V_1$ and $V_2$ for the pp-wave with $1/R^2$ terms:
\begin{eqnarray}
&& V_1 = \frac{\partial \phi}{\partial x_1} V_{\phi} 
+ \frac{\partial r}{\partial x_1}V_r 
= - \frac{\sin\phi}{r}V_{\phi} 
= - \frac{x'_2 + r_0^2}{x_1^2 + (x'_2+r_0)^2}V_{\phi}  \nn \\ 
&& \quad\,\, = \frac{1}{2\sqrt{x'_2{}^2 + y^2}} - \frac{1}{R^2}\left[
\frac{1}{2} + \frac{x'_2(x_1^2+x'_2{}^2 + y^2)}{4(x'_2{}^2 + y^2)^{3/2}}
\right] + \mathcal{O}(1/R^4) \label{v1} \\ 
&& \quad\,\, 
= \frac{1}{r_1^2 + r_2^2} - \frac{r_1^2}{R^2(r_1^2 + r_2^2)} 
+ \mathcal{O}(1/R^4)\,, \nn  \\  
&& V_2 = \frac{\partial \phi}{\partial x_2} V_{\phi} 
+ \frac{\partial r}{\partial x_2}V_r = \frac{\cos\phi}{r}V_{\phi} 
= \frac{x_1}{x_1^2 + (x'_2 + r_0)^2}V_{\phi} \nn \\ 
&& \quad\,\, = - \frac{x_1}{2R^2\sqrt{x'_2{}^2 + y^2}} + \mathcal{O}(1/R^4) 
= - \frac{x_1}{R^2(r_1^2 +r_2^2)} + \mathcal{O}(1/R^4)\,.  \label{v2} 
\end{eqnarray} 
The above $V_1$ and $V_2$ satisfy the differential equations (\ref{z-v})
with the function $z$ given in (\ref{z})\,. As a remark, the constant
term $1/2R^2$ in the expression of $V_1$ (\ref{v1}) would not be
determined by solving the differential equation but it is properly 
determined by carefully considering the Penrose limit.
It is also worth remaking that the relation $z = x_2' V_1 - x_1 V_2 $, mentioned in \cite{Horava}, begins to fail at the $1/R^2$-order:
\begin{eqnarray*}
x_2' V_1 - x_1 V_2
&=&
	\frac{{x_2'}}{2 {\sqrt{y^2 + {{x_2'}}^2}}}
	 - \frac{1}{4 R^2}
	\Bigg[{2 {x_2'} + \frac{{{x_2'}}^2 \left( y^2 + {{x_2'}}^2 \right)  - 
        {{x_1}}^2 \left( 2 y^2 + {{x_2'}}^2 \right) }{{\left( y^2 + {{x_2'}}^2 \right) }^{\frac{3}{2}}}}\Bigg]
 +{\cal O}(1/R^4)\,.
\end{eqnarray*}

\par
By putting the functions (\ref{z}), (\ref{v1}) and (\ref{v2}) into the
metric (\ref{metric}), we can derive the metric:
\begin{eqnarray}
&& ds^2 = -2dtdx_1 -(r_1^2 +r_2^2) dt^2 + d\vec{r_1}^2 + d\vec{r_2}^2  \\ 
&& \qquad \qquad + \frac{1}{R^2}
\Bigl[dx_1^2 + \frac{1}{2}(r_1^2 d\vec{r_1}^2 - r_2^2 d\vec{r_2}^2) 
+ \frac{1}{2}(r_2^4 - r_1^4) dt^2 \nn \\ 
&& \qquad \qquad + dt dx_1 (r_1^2 + r_2^2) 
+ 2x_1(r_1dr_1dt - r_2dr_2 dt) 
\Bigr] + \mathcal{O}(1/R^4)\,, \nn
\end{eqnarray}
where we have written down the metric in terms of the coordinates 
$(r_1,r_2)$ rather than $(y,x_2)$\,. We also used the following expansions
of $y\e^{G}$ and $y\e^{-G}$\,: 
\begin{eqnarray}
&& y\e^{G} = 
\frac{y^2}{-x'_2 + \sqrt{x'_2{}^2 + y^2}} 
+ \frac{y^2(x_1^2 +x'_2{}^2 + y^2)}{2R^2(x'_2{}^2 + y^2 -x'_2
\sqrt{x'_2{}^2 +y^2})} 
+ \mathcal{O}(1/R^4)  \nn \\ 
&& 
\qquad = r_1^2 + \frac{r_1^4}{2R^2} + \mathcal{O}(1/R^4)\,, \nn \\
&& 
y\e^{-G} = 
-x'_2 + \sqrt{x'_2{}^2 + y^2} 
- \frac{(x_1^2+x'_2{}^2 + y^2)
\left(x'_2{}^2+y^2  -x'_2\sqrt{x'_2{}^2+y^2}\right)}{2R^2 (x'_2{}^2 + y^2)} 
+ \mathcal{O}(1/R^4)  \nn \\ 
&& 
\qquad\,\,\, = r_2^2 - \frac{r_2^4}{2R^2} + \mathcal{O}(1/R^4)\,.  \nn 
\end{eqnarray}
Furthermore performing the shift of $x_1$\, as $ x_1 = x'_1 + x'_1 x_2'/R^2\,, $ and identifying as $t \equiv x^+$ and $x'_1 \equiv - x^-$ gives the pp-wave metric with $1/R^2$ corrections considered in
\cite{Schwarz}:
\begin{eqnarray}
&& \hspace*{-0.5cm}
ds^2 = 2 dx^+dx^- + d\vec{r_1}^2 + d\vec{r_2}^2 - (r_1^2 +r_2^2)(dx^+)^2  \\ 
&& \quad~ 
+ \frac{1}{R^2}\left[-2y^2dx^-dx^+ + \frac{1}{2}(r_2^4-r_1^4)(dx^+)^2 
+ (dx^-)^2 
+ \frac{1}{2}r_1^2d\vec{r_1}^2 - \frac{1}{2}r_2^2d\vec{r_2}^2\right] 
+ \mathcal{O}\left(1/R^4\right)\,.    \nn 
\end{eqnarray}
Here the convention of the light-cone coordinates is absorbed into the
identification between $x^-$ and $x'_1$\,. As an additional remark, the
shift of $x_1$ does not change the expression of $z$, $V_1$, and
$V_2$ at the order of $1/R^2$\,.

\vspace*{0.3cm}

Finally we should note that the function $z$ including the $1/R^2$-corrections has the same boundary condition at $y=0$ as in the case of $z$ {\it without} $1/R^2$-corrections, although the $V_2$  becomes non-zero due to the $1/R^2$-correction.
That is,  $1/R^2$-corrections are irrelevant to the droplet.

\section{Higher Order $1/R^2$-Corrections}
In this section we further investigate higher order $1/R^2$-corrections at the order of $1/R^6$\,.
Using the relation (\ref{expan}) found in the section 3, the higher order $1/R^2$-corrections of the $z$ can be computed as 
\begin{eqnarray}
z(x_1,x_2',y)
&=&
	\frac{x_2^\prime}{2\,{\sqrt{{x_2^\prime}^2 + y^2}}} + 
  \frac{y^2\,\left( {x_2^\prime}^2 + y^2 + {{x_1}}^2 \right) }
   {4\,R^2\,{\left( {x_2^\prime}^2 + y^2 \right) }^{\frac{3}{2}}} - 
  \frac{3\,x_2^\prime\,y^2\,{\left( {x_2^\prime}^2 + y^2 + {{x_1}}^2 \right) }^2}
   {16\,R^4\,{\left( {x_2^\prime}^2 + y^2 \right) }^{\frac{5}{2}}} \nonumber\\
&&\quad
	-\frac{ y^2\,\left( y^2 - 4\,x_2^{\prime 2} \right) \,
      {\left( y^2 + {{x_1}}^2 + x_2^{\prime 2} \right) }^3  }{32\,R^6\,
    {\left( y^2 + x_2^{\prime 2} \right) }^{\frac{7}{2}}}
	+{\cal O}(1/R^8). \label{z6-1}
\end{eqnarray}
We can explicitly check that the function $z$ obeys the differential equation (2.3).
It is worth while noting that the droplet configuration $z(x_1,x_2',y=0)$ of (\ref{z6-1}) is the same as the pp-wave one (the right hand side in Fig.\,\ref{pp-wave:fig}) even at the order of $1/R^6$.
Hence it is expected that the droplet configuration would be the same as the pp-wave one even if we include {\it any finite} $1/R^2$-corrections.
As a matter of course, when we include all order $1/R^2$-corrections, the droplet configuration would become the AdS$_5$$\times$S$^5$ one (the left hand side in Fig.\,\ref{pp-wave:fig}).
One possible reason why the different geometries give the same droplet configuration is that the size of the droplet for the pp-wave is infinite in comparison to the finite size droplet for the AdS case.
Hence it might be necessary to give more boundary conditions at infinity as in the argument given in the case of pp-wave (with no corrections) \cite{LLM}, when we consider the $1/R^2$-corrections. 
Another possible explanation is that the two limits $ R \to \infty$ and $ y \to 0$ do not commute.
When we first take the limit $R\to \infty$, the droplet boundary goes to infinity.
This limit would hide the behavior of $z$ near the boundaries at infinity.
Then the limit of $ y \to 0$ would give the same droplet configuration as the one without $1/R^2$-corrections.

In order to obtain $z$ as a function of $(r_1,r_2,x_1)$ we expand $y$ and $x_2'$ in terms of $r_1$ and $r_2$,
\begin{eqnarray*}
y
&=&
	 r_1 r_2 +\frac{r_1r_2}{4R^2}(r_1^2-r_2^2)
	+\frac{{r_1}{r_2}}{16 R^4} \left( r_1^4 - r_1^2 r_2^2 + r_2^4 \right)
	+\frac{r_1 r_2}{64 R^6}\left(r_1^2 - r_2^2 \right)\left( r_1^4 + r_2^4 \right)
	+{\cal O}(1/R^8),
\\
x_2'
&=&
	 \frac{1}{2}(r_1^2-r_2^2)+\frac{1}{8R^2}[(r_1^2-r_2^2)^2- 4 x_1^2]
	+\frac{r_1^2-r_2^2}{32 R^4} \left({{r_1}}^4 - {{r_1}}^2 {{r_2}}^2 + {{r_2}}^4 + 8 {{x_1}}^2 \right) \\
&&	+\frac{{\left( {{r_1}}^2 - {{r_2}}^2 \right) }^2 \left( {{r_1}}^4 + {{r_2}}^4 \right)  - 
    8 {\left( {{r_1}}^2 - {{r_2}}^2 \right) }^2 {{x_1}}^2 - 16 {{x_1}}^4}{128 R^6}
+{\cal O}(1/R^8).
\end{eqnarray*}
By using these relations, we can rewrite (\ref{z6-1}) in terms of $r_1$ and $r_2$ as follows:
\begin{equation}
z=
	\frac{r_1^2-r_2^2}{2(r_1^2+r_2^2)}+\frac{r_1^2 r_2^2}{2R^2(r_1^2+r_2^2)}
	+\frac{r_1^2 r_2^4 - r_1^4 r_2^2   }
  {16 R^4 \left(r_1^2 + r_2^2 \right) }
	-\frac{ r_1^4 r_2^4  }{32 R^6 \left(r_1^2 + r_2^2 \right) }
	+{\cal O}(1/R^8). \label{z6-2}
\end{equation}

\par
In order to obtain the metric including the $1/R^6$-corrections we compute $V_1$, $V_2$, $ye^G$ and $ye^{-G}$ in terms of $x_1$, $r_1$ and $r_2$.
The results for $V_1$ and $V_2$ are
\begin{eqnarray}
V_1 &=&
	\frac{1}{2 {\sqrt{y^2 + x_2^{\prime 2}}}}
  - \frac{{x_2'} \left( y^2 + x_1^2 + {x_2'}^2 \right) }
      {4 { ( y^2 + {{x_2'}}^2 ) }^{{3}/{2}} R^2}
  -\frac{1}{2R^2}  +\frac{x_2'}{2 R^4} \label{V1R6}\\
&&
	+\frac{1}{16 R^4 
    {({x_2^\prime}^2 + y^2)}^{5/2}}
	\left[{{ ({x_2^\prime}^2 + y^2 ) }^2 
     ( 6 {x_2^\prime}^2 + 3 y^2 )
	 - 6 y^2 ( {x_2^\prime}^2 + y^2 )
	  {{x_1}}^2 + ( 2 {x_2^\prime}^2 - y^2 )  x_1^4} \right] \nonumber\\
&&
	+\frac{1}{32 R^6 ( y^2 + {x_2'}^2 )^4}
	\bigg\{-x_2' \sqrt{y^2 + x_2^{\prime 2}}
     \bigg[ x_1^4 ( y^2 + {x_2'}^2 ) ( -13 y^2 + 2 {{x_2'}}^2)
 	  +x_1^6 ( -3 y^2 + 2 {x_2'}^2 ) \nonumber\\
&&\quad
	  - 3 x_1^2 ( y^2 + {{x_2'}}^2 )^2 
        ( 11 y^2 + 6 {x_2'}^2 )
	  +( y^2 + {x_2'}^2 )^3 ( 9 y^2 + 14 {x_2'}^2 ) \bigg]\bigg\}
	+\frac{x_1^2-{x_2'}^2}{2 R^6}
	+{\cal O}(1/R^8)
  \nonumber\\
&=& 
	\frac{1}{r_1^2 + r_2^2} - \frac{r_1^2}{R^2 ( r_1^2 + r_2^2 ) }
	-\frac{ r_2^4  + r_1^2 r_2^2 -7 r_1^4  + 8 x_1^2  }
  {16 R^4 ( r_1^2 + r_2^2 ) }
	+\frac{r_1^2(16 x_1^2 -2 r_1^4 + r_2^4) -r_2^2(8x_1^2-r_1^4)}
	{16 R^6 ( r_1^2 + r_2^2 ) }  \nonumber\\
&& +{\cal O}(1/R^8),  \nonumber
\end{eqnarray}
\begin{eqnarray}
V_2
&=&
	-\frac{{x_1}}{2 R^2 {\sqrt{y^2 + x_2^{\prime 2}}}}
	+\frac{x_1 x_2^\prime}{4 R^4}
	\frac{ x_1^2 + 3 ({x_2^\prime}^2 + y^2)}{{(x_2^{\prime 2} + y^2) }^{{3}/{2}}} 
	+\frac{x_1}{2 R^4} -\frac{x_2'}{R^6} \label{V2R6} \\
&&\quad
	 +\frac{ x_1 \left\{  x_1^4 ( y^2 - 2 x_2^{\prime 2})  
	 - 3 ( y^2 + x_2^{\prime 2} )^2 ( y^2 + 6 x_2^{\prime 2})
	  + x_1^2 ( 6 y^4 + 2 y^2 x_2^{\prime 2} - 4 x_2^{\prime 4} )  \right\}}
	{16 R^6 {\left( y^2 + {{x_2'}}^2 \right) }^{\frac{5}{2}}}
	+{\cal O}(1/R^8)
	 \nonumber\\
&=&
	-\frac{{x_1}}{R^2 \left( {{r_1}}^2 + {{r_2}}^2 \right) }
	+\frac{\left( 3 {{r_1}}^2 - {{r_2}}^2 \right)  {x_1}}
	  {2 R^4 \left( {{r_1}}^2 + {{r_2}}^2 \right) }
	-\frac{ \left( 17 {{r_1}}^4 - 13 {{r_1}}^2 {{r_2}}^2 + {{r_2}}^4 \right)  {x_1}
      }{16 R^6 \left( {{r_1}}^2 + {{r_2}}^2 \right) }
	+{\cal O}(1/R^8). \nonumber
\end{eqnarray}
We can check that $V_1$ and $V_2$ obey the equations (\ref{z-v}).
It is worth noting again that the $y$-independent terms in (\ref{V1R6}) and (\ref{V2R6}) would not be completely determined from the differential equations (\ref{z-v}), although they constrain the relation between the undetermined terms in $V_1$ and the ones in $V_2$. 

\par
Using Eq. (\ref{z6-2}) the results for $ye^G$ and $y e^{-G}$ are
\begin{eqnarray*}
y e^G
&=&
	r_1^2
	 + \frac{r_1^4}{2 R^2}
 	+\frac{3 r_1^6}{16 R^4}
	+\frac{r_1^8}{16 R^6}
	+{\cal O}(1/R^8),
\\
y e^{-G} 
&=&
	r_2^2 - \frac{r_2^4}{2 R^2} + \frac{3 r_2^6}{16 R^4}
	-\frac{r_2^8}{16 R^6} +{\cal O}(1/R^8).
\end{eqnarray*}

\par
Thus the resulting metric is
\begin{eqnarray*}
ds^2
&=&
	 2dx^+ dx^- -(r_1^2+r_2^2)(dx^+)^2 +d\vec{r}_1{}^2 +d\vec{r}_2{}^2 \nonumber\\
&&
 +\frac{1}{R^2}\bigg[
   2 r_2^2 dx^+ dx^- + (dx^-)^2 - (r_1^4 - r_2^4)(dx^+)^2
 + \frac{1}{2}r_1^2d\vec{r}_1{}^2 -\frac{1}{2} r_2^2 d\vec{r}_2{}^2 \bigg] \nonumber\\
&&
	+\frac{1}{R^4}
	\bigg[ - r_2^4 dx^+ dx^-
	 - r_2^2 (dx^-)^2
	 -\frac{3}{16} ( r_1^6 + r_2^6 ) (dx^+)^2
	+ \frac{3}{16} r_1^4  {{{{d\vec{r}}}_1}}^2   + \frac{3}{16} r_2^4 {{{{d\vec{r}}}_2}}^2  \bigg] \nonumber\\
&&
+\frac{1}{16 R^6}
	\bigg[ 6 r_2^6 dx^+ dx^- +8r_2^4 (dx^-)^2
	-( r_1^8 - r_2^8 ) (dx^+)^2
	+ r_1^6 {{{{d\vec{r}}}_1}}^2   - r_2^6 {{{{d\vec{r}}}_2}}^2 \bigg]
	+{\cal O}(1/R^8) 
\end{eqnarray*}
where we have used the coordinate-transformation from  $x_1$  to $x_1'$ defined as
\begin{eqnarray*}
x_1
 &\equiv&
	x_1'+\frac{x_1' x_2'}{R^2} +\frac{x_1'{}^3}{2 R^4}
	 +\frac{x_1'{}^3 x_2'}{3 R^6}
	+{\cal O}(1/R^8)
 \nonumber \\
 &=& x_1'+ \frac{x_1'(r_1^2-r_2^2)}{2R^2}
	-\frac{x_1^{\prime 3}}{6 R^4} + \frac{x_1'}{8 R^4} (r_1^2 -r_2^2)^2 
 -\frac{{x_1'}^3 ( r_1^2 - r_2^2 ) }{12 R^6} + 
  \frac{x_1'(r_1^2-r_2^2)}{32 R^6}
	\left[r_1^4 -r_1^2 r_2^2 +r_2^4 \right] \\
&& \quad
		+{\cal O}(1/R^8),
\end{eqnarray*}
and the same identifications in section 3 : $t \equiv x^+$ and $x_1' \equiv -x^-$ \,.

\section{Concentric Rings}

\begin{figure}[hbtp]
 \begin{tabular}{cc}
\hspace*{-7mm}
  \begin{minipage}{0.5\hsize}
   \begin{center}
    \resizebox{65mm}{!}{\includegraphics{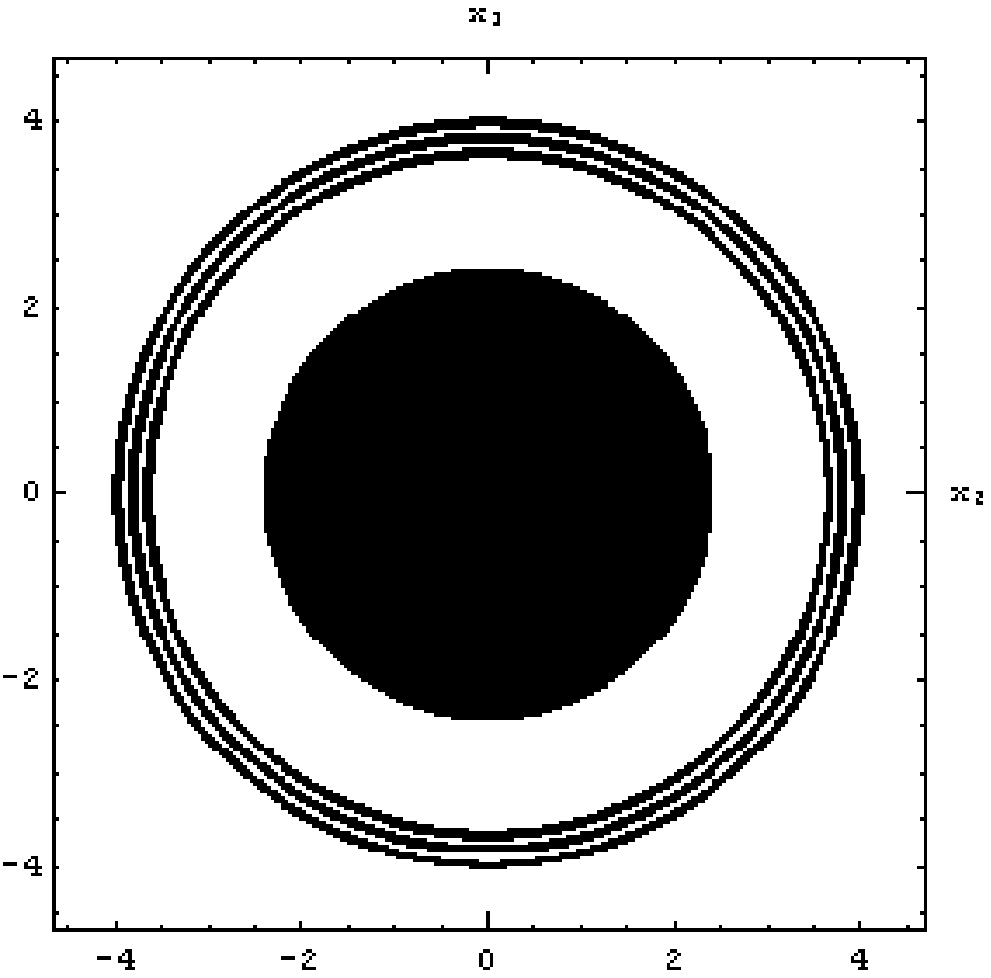}}
   \end{center}
   \vspace*{-9mm}
   \caption{Configuration of concentric rings.}\label{con-ring:fig}
  \end{minipage}
  \begin{minipage}{0.5\hsize}
   \begin{center}
    \resizebox{65mm}{!}{\includegraphics{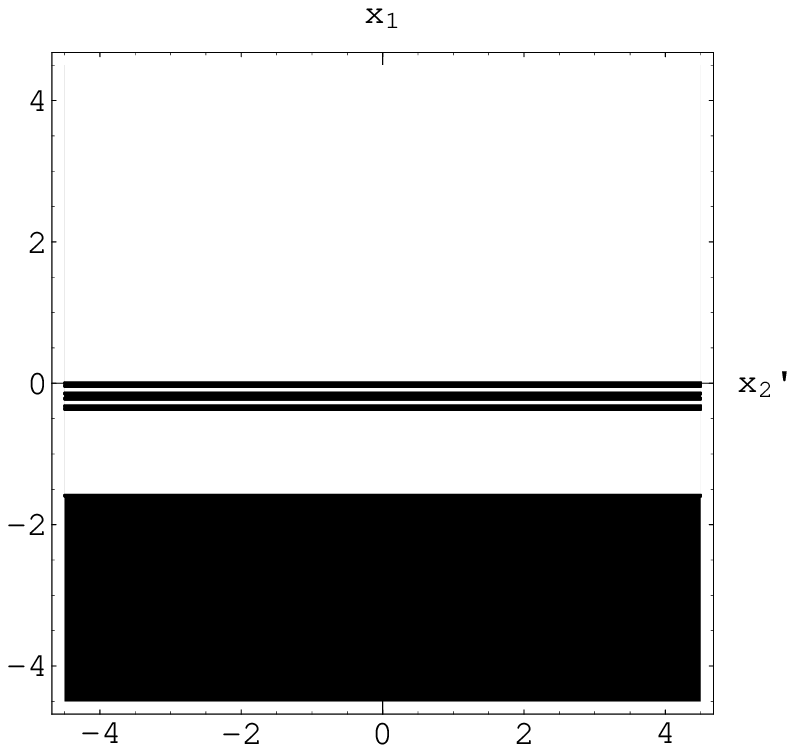}}
   \end{center}
   \vspace*{-9mm} 
   \caption{Penrose limit of concentric rings.}\label{con-pp:fig}
  \end{minipage}
 \end{tabular}
\end{figure}

As a simple extension of the discussion in section \ref{sec3}, let us
consider the geometry characterized by a family of concentric rings
\cite{LLM} (see Fig.\,\ref{con-ring:fig}). This solution is given by the
following $z,V_r,V_\phi$:
\begin{eqnarray}
&& \tilde{z} = \frac{1}{2}\sum_{n=1}^N(-1)^{n+1}\left(
\frac{r^2-r^2_n+y^2}{\sqrt{(r^2+r^2_n+y^2)^2 -4r^2r^2_n}} -1
\right)\,, \nn \\
&& V_r = 0\,, \quad V_{\phi} = \frac{1}{2}\sum_{n=1}^N (-1)^n 
\left(\frac{r^2+r^2_n +y^2}{\sqrt{(r^2 + r^2_n +y^2)^2 -4 r^2 r^2_n}} 
-1 \right)\,, \nn 
\end{eqnarray}
where we have used the polar coordinates $(r,\phi)$ instead of
$(x_1,x_2)$\,. The $r_1$ is the radius of the outermost circle, $r_2$
the next one and so on.  This background is time-independent and in
certain limits can be thought of as a configuration of smeared S$^5$
giants and/or their AdS$_5$ duals. 

It is easy to apply the previous analysis to this case. All we have to
do is to introduce the shifts of variables as follows:
\begin{eqnarray}
x_2 - r_n = (x_2 - r_0) - (r_n - r_0) \equiv x'_2 - x_2^{(n)} \,.  
\end{eqnarray}
Then the radius coordinate $r$ is expanded as 
\begin{eqnarray}
r = \sqrt{x_1^2 + (x_2'- x_2^{(n)} + r_n)^2} = (x_2'-x_2^{(n)} +r_n) + 
\frac{x_1^2}{2(x_2'-x_2^{(n)} + r_n)} + \cdots\,.
\end{eqnarray} 
In addition we assume that $r_0$ is much bigger than $x_2^{(n)}$ (thin
ring approximation) and expand $r_n$ as
\[
 r_n = r_0\left(1 + \frac{x_2^{(n)}}{r_0}\right) \equiv r_0 
\left(1 + \epsilon^{(n)} \right)\,. 
\]
The remaining part of the analysis is similar to that in the
AdS$_5\times$S$^5$ ($N=1$ case) and the $1/R^2$ corrections in this case
can be also evaluated. The resulting function $z$ after taking the
Penrose limit in the configuration of the concentric rings is
\begin{eqnarray}
z = \frac{1}{2}\sum_{n=1}^N (-1)^{n+1}\left\{
\frac{x'_2 - x_2^{(n)}}{\sqrt{(x'_2-x_2^{(n)})^2 + y^2}}
+ \frac{y^2\left[(x'_2-x_2^{(n)})^2 
+ x_1^2 + y^2 \right]}{2R^2\left[(x'_2-x_2^{(n)})^2 +y^2 \right]^{3/2}} 
\right\}
+ \mathcal{O}(1/R^4)\,. \label{z:strip}
\end{eqnarray}  
Then $V_1$ and $V_2$ are given by 
\begin{eqnarray}
&& V_1 = \sum_{n=1}^N (-1)^{n-1}\Biggl\{
\frac{1}{2\sqrt{(x'_2-x_2^{(n)})^2 + y^2}}  \\ 
&& \hspace*{3.3cm} - \frac{1}{R^2}\Biggl[
 \frac{1}{2} + \frac{(x'_2 - x_2^{(n)})\bigl[x_1^2 + (x'_2-x_2^{(n)})^2 +y^2
\bigr]}{4\bigl[(x'_2-x_2^{(n)})^2 + y^2\bigr]^{3/2}}
\Biggr]
\Biggr\} + \mathcal{O}(1/R^4)\,, \nn \\ 
&& V_2 = \sum_{n=1}^N (-1)^{n}\frac{x_1}{2R^2\sqrt{(x'_2-x_2^{(n)})^2 +y^2}} 
+ \mathcal{O}(1/R^4)\,. 
\end{eqnarray}
The droplet configurations at $y=0$ are a set of stripes as noted by LLM
\cite{LLM} (see Fig.\,\ref{con-pp:fig}). The leading part of the above
result agrees with the one in \cite{EM}. We have plotted two graphs
(Figs.\,\ref{con-ring:fig} and \ref{con-pp:fig}) by using the
contour plot in the Mathematica with the data: $r_1 = R^2,~r_2 = 0.99
R^2,~r_3 = 0.96 R^2,~r_4=0.95 R^2, ~r_5=0.92 R^2,~r_6=0.91 R^2,~r_7 =
0.6 R^2,~R=2$\,. Our result (\ref{z:strip}) surely reproduces a set of
strips from the concentric rings via the Penrose limit.

\par
In a similar way we can compute the higher order corrections for concentric rings.

\section{A Conclusion and Discussions}

We have discussed how to take a Penrose limit in bubbling 1/2 BPS
geometries at the stage of a single function $z(x_1,x_2,y)$\,.  Taking
the Penrose limit for the function $z$ for the AdS, we can directly
obtain the $z$ for the pp-wave with $1/R^2$-corrections. It satisfies
the differential equation and leads to the pp-wave metric with
$1/R^2$-corrections. In particular, our result reproduces the pp-wave
metric used in \cite{Schwarz}.

It should be however noted that the function with $1/R^2$-corrections
has the same boundary condition at $y=0$ as the $z$ without the
corrections.  Hence the $1/R^2$-corrections are not determined by
naively imposing a boundary condition at $y=0$\,, and it would be necessary
to take more careful treatments. But, by considering the Penrose limit at the stage of the single function $z$ for the AdS space (or metrics of other spacetimes), it is possible to properly take $1/R^2$-corrections into account in bubbling 1/2 BPS geometries.

We also considered the higher $1/R^2$-contributions to the $z$\,.
We saw that higher-order corrections do not modify the half-filling configuration as well as the second order corrections.
That is, the $z$ has the same boundary condition at $y=0$ as the $z$ with no correction. 
One possible reason why the different geometries give the same droplet configuration is that the size of the droplet for the pp-wave is infinite in comparison to the finite size droplet for the AdS case.
It might be necessary to give more boundary conditions at infinity as in the argument given in the case of pp-wave (with no corrections) \cite{LLM}, when we consider the $1/R^2$-corrections. 
It was also found that the $y$-independent terms in $V_1$ and  $V_2$ would not be completely determined from the differential equations (\ref{z-v}), 
although they constrain the relation between the undetermined terms in $V_1$ and the ones in $V_2$\,.
It would be nice to apply our discussion to other metrics (for example, \cite{other-dim}) and derive the corresponding function $z$ (with $1/R^2$-corrections). 
On the other hand, it is interesting to consider the description of $1/R^2$-corrections in terms of the free fermion in the SYM side.
In this direction it would be valuable to comment on the work of Horava and Shepard \cite{Horava}.
They also considered the Penrose limit of LLM geometries, and in particular, showed that the Penrose limit towards the (nearly) singular null geodesic of the geometry (nearly) at topological transition is equivalent to the well-known double-scaling limit of the matrix model that defines two-dimensional noncritical string theory (in fact, the Type 0 version of it \cite{0B}, since both sides of the Fermi sea are filled). 
Hence the $1/R^2$-corrections in the Penrose limit would correspond to the corrections to the double scaled matrix model.
It is also worthwhile to say that the non-relativistic free fermions would become relativistic in the Penrose limit according to the LLM's observation \cite{LLM}.
Then it would be expected that the $1/R^2$-corrections interpolate between the non-relativistic fermions and relativistic ones in the Penrose limit.

\section*{Acknowledgments}
We would like to thank H.~Fuji, Y.~Susaki, A.~Tsuchiya and A.~Yamaguchi for useful
discussion. The work of K.~Y.\ is supported in part by JSPS Research
Fellowships for Young Scientists.

\end{document}